\documentclass[prb,preprint,showpacs,floatfix]{revtex4}

\usepackage{amsmath}
\usepackage{epsfig,amssymb,txfonts} 
\usepackage[hang,raggedright]{subfigure}
\usepackage{multirow}
\usepackage{amssymb}
\usepackage{amsfonts}

\newcommand*{\be}[0]{\begin{equation}}
\newcommand*{\ee}[0]{\end{equation}}
\newcommand*{\beu}[0]{\begin{equation*}}
\newcommand*{\eeu}[0]{\end{equation*}}
\newcommand*{\ba}[0]{\begin{array}}
\newcommand*{\ea}[0]{\end{array}}
\newcommand*{\bfig}[0]{\begin{figure}[!ht]}
\newcommand*{\efig}[0]{\end{figure}}
\newcommand*{\bfigwide}[0]{\begin{figure*}}
\newcommand*{\efigwide}[0]{\end{figure*}}
\newcommand{\CuAg}{$\phi_{\text{CuAg}}(r)$}
\newcommand{\CuCu}{$\phi_{\text{CuCu}}(r)$}
\newcommand{\AgAg}{$\phi_{\text{AgAg}}(r)$}

\newlength{\wholefigwidth}
\newlength{\smallfigwidth}
\newlength{\halfsmallfigwidth}
\newcommand{\Fig}[1]{Fig.~\ref{fig:#1}}
\newcommand{\Tab}[1]{Table~\ref{tab:#1}}
\newcommand{\Sec}[1]{Section~\ref{sec:#1}}
\newcommand{\Eqn}[1]{Eqn.~\ref{eqn:#1}}

\begin{document}

\title{Cu/Ag EAM Potential Optimized for Heteroepitaxial Diffusion from
\textit{ab initio} Data}

\author{Henry H. Wu}
\author{Dallas R. Trinkle}
\affiliation{Department of Materials Science and Engineering, University of
Illinois, Urbana-Champaign}

\begin{abstract}
A binary embedded-atom method (EAM) potential is optimized for Cu on
Ag(111) by fitting to \textit{ab initio} data.  The fitting database
consists of DFT calculations of Cu monomers and dimers on Ag(111),
specifically their relative energies, adatom heights, and dimer
separations.  We start from the Mishin Cu-Ag EAM potential and first modify
the Cu-Ag pair potential to match the FCC/HCP site energy difference then
include Cu-Cu pair potential optimization for the entire database.  The
optimized EAM potential reproduce DFT monomer and dimer relative energies
and geometries correctly.  In trimer calculations, the potential produces
the DFT relative energy between FCC and HCP trimers, though a different
ground state is predicted.  We use the optimized potential to calculate
diffusion barriers for Cu monomers, dimers, and trimers.  The predicted
monomer barrier is the same as DFT, while experimental barriers for
monomers and dimers are both lower than predicted here.  We attribute the
difference with experiment to the overestimation of surface adsorption
energies by DFT and a simple correction is presented.  Our results show
that the optimized Cu-Ag EAM can be applied in the study of larger Cu
islands on Ag(111).
\end{abstract}

\maketitle

\section{Introduction}
Knowledge of the surface diffusion dynamics for small atom clusters is
critical to understanding heteroepitaxial thin film growth.  While numerous
experiments%
\cite{Wen:1994uq,Kellogg:1991yq,Bartelt:1999kx,Antczak:2007fj}
and computer simulations%
\cite{Papanicolaou:1998uq,Montalenti:1999fj,Bogicevic:1999lr,Lorensen:1999qy}
have studied homogeneous systems, less is known about lattice mismatched
heterogeneous systems%
\cite{Chirita:1999fk,Papathanakos:2002qy,Brune:1999fk}
and their interesting diffusion kinetics.  In this study, we consider Cu on
Ag(111),\cite{Morgenstern:2004uq} a system with a lattice mismatch of
12\%.\cite{Ozolins:1998lr} The lattice mismatch induces strain in both the
island and substrate and has been predicted to promote rapid
diffusion.\cite{Hamilton:1996lr}

To accurately compute the energetics of surface island systems, first
principle density-functional theory (DFT) calculations are preferred to
empirical potentials.  However, DFT methods are too computationally
intensive to efficiently search the phase space of each island and accurate
classical potentials are needed to characterize island diffusional
dynamics.  The embedded atom method\cite{Foiles:1986fj} (EAM) is well
suited for metallic systems combining pair interactions with an atomic
embedding energy term dependent on the local ``electron density.''
\Tab{summary} shows that other EAM potentials were unable to reproduce DFT
calculated Cu island energies and geometries on Ag(111), motivating the
search for a new potential.

We optimize a new EAM potential for Cu on Ag(111) using monomer and dimer
data.  \Sec{method} explains the DFT and EAM calculation parameters in
detail.  \Sec{optim} presents the procedure for the potential optimization.
The energetics and diffusion results from the new EAM for monomers, dimers,
and trimers are reported in \Sec{results}.  We justify the new potential
for the study of small Cu islands on Ag(111) surface by comparing the
calculated results to experimental and DFT values in \Sec{discussion}.

\section{Computational Details}
\label{sec:method}
The density-functional theory calculations are performed with
\textsc{vasp},\cite{Kresse93,Kresse96b} a density-functional code using a
plane-wave basis and ultrasoft Vanderbilt-type
pseudopotentials.\cite{Vanderbilt90,Kresse94}  The local-density
approximation as parametrized by Perdew and Zunger\cite{Perdew1981} and a
plane-wave kinetic-energy cutoff of 200~eV ensures accurate treatment of
the Cu and Ag potential.  We treat the $s$ and $d$ states as valence,
corresponding to an Ar and Kr core atomic reference configuration for Cu
and Ag, respectively.  The $(111)$ surface slab calculations used a
$3\times3$ geometry with 6 $(111)$ planes of Ag and 6 $(111)$ planes of
vacuum; the $k$-point meshes for the surface slab calculations are
$8\times8\times1$, with a Methfessel-Paxton smearing of 0.25~eV.

EAM energy values were computed with the \textsc{lammps} molecular dynamics
package.\cite{Plimpton:1995lr} The monomer and dimer results in
\Tab{summary} are obtained using a periodic 3$\times$3 cell of 6 (111)
planes.  The trimers are calculated with 4$\times$4 periodic cells.
Results presented in section IV are from 6$\times$6 periodic cells, where
our potential predicts a finite-size effect of less than 5meV compared to
the 3$\times$3 cell.  Transition energy barriers are determined with nudged
elastic band\cite{Mills:1994yq} calculations after initial and final states
have been found through molecular dynamics or the dimer search
method.\cite{Henkelman:1999lr} Attempt frequency prefactors are computed
with the Vineyard formula,\cite{Vineyard:1957vn} taking the ratio between
the product of harmonic vibrational frequencies at the initial state and
the saddle point.

\section{Optimization Procedure}
\label{sec:optim}
In EAM, the total energy of the system is given by
\beu
\begin{split}
E_{\text{tot}} &= \frac{1}{2}\sum_{ij}\phi_{ij}(r_{ij}) +
\sum_{i}F_{i}(\rho_{i, \text{tot}})\\
\rho_{i, \text{tot}} &= \sum_{j\neq i}\rho_{j}(r_{ij}),
\end{split}
\eeu
where $\phi_{ij}(r_{ij})$ is the pair potential interaction between atoms
{\it i} and {\it j} separated by a distance of $r_{ij}$ and $F_{i}(\rho_{i,
\text{tot}})$ is embedding energy of atom {\it i} in the superposition of
atomic electron densities $\rho_{j}(r_{ij})$.  The Mishin CuAg binary EAM
potential\cite{Mishin:2001ym,Williams:2006qy} is described by seven
functions: \CuCu, \AgAg, \CuAg, $\rho_{\text{Cu}}(r)$,
$\rho_{\text{Ag}}(r)$, $F_{\text{Cu}}(\rho)$, and $F_{\text{Ag}}(\rho)$.
The Mishin EAM embedding energy functions and electron density functions
are not changed in our optimization.  Only the Cu-Ag and Cu-Cu pair
potentials are modified to fit our DFT optimization database.  We forgo
modification of the Ag-Ag potential because the distance between relaxed
EAM Ag(111) planes are within 3\% of the relaxed DFT Ag surface.

Cu monomers and dimers are building blocks for larger islands, making them
ideal choices for the optimization database.  The Ag(111) surface is
divided into FCC and HCP sites, depending on the atomic configuration
continuing from the top two layers of Ag.  For monomers, the single Cu atom
rests at either a FCC or HCP site.  For dimers, four different
configuration of the Cu pair can be formed, FCC-FCC (FF), HCP-HCP (HH), and
two types of FCC-HCP (FH$_{\text{short}}$ and FH$_{\text{long}}$).  The two
FH dimers (c.f.~\Fig{dimer}) are differentiated by their neighboring Ag atoms,
the two triangles of Ag neighbors can share a side (FH$_{\text{short}}$),
or share a corner (FH$_{\text{long}}$).  The optimization database consists
of the relative DFT energies between FCC and HCP monomers and all four
dimers, geometric information on the heights of monomers and dimers above
the Ag surface, and the Cu-Cu separation length.  We minimize the total
root-mean-square (rms) error of the energy differences and balance that
with the total rms error of the heights and lengths.

\bfig
\includegraphics[width=\smallfigwidth]{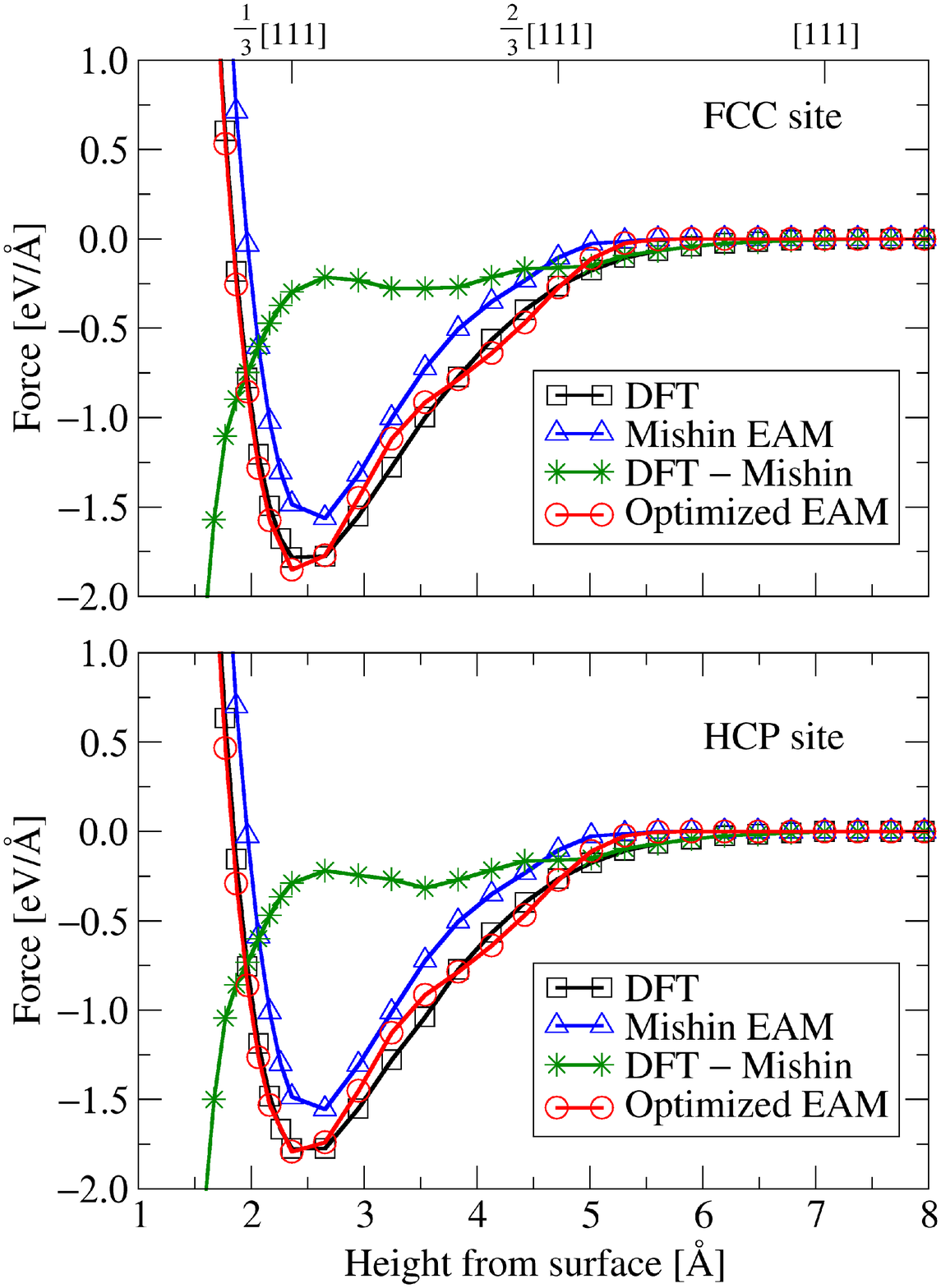}
\caption{Force on a Cu atom above an FCC site or an HCP site on the
unrelaxed Ag(111) surface calculated by DFT, Mishin EAM, and the new
optimized EAM.  The ultrasoft pseudopotential DFT force displays a stronger
and wider interaction between the Cu atom and the Ag surface than Mishin
EAM.  The average force difference (FCC and HCP) between DFT and Mishin EAM
is used to produce the force-matched \CuAg\ in \Fig{CuAg}.  The final
optimized EAM potential deviates between 3\AA--5\AA\ with a maximum
deviation of 0.15eV/\AA\,with respect to the DFT calculated forces.}
\label{fig:111z}
\efig

In \Fig{111z} the DFT force of a Cu atom evaporating from a perfect
(unrelaxed) Ag(111) surface is plotted, and is used in addition to the
database.  Starting from a height of $\frac{1}{6}$[111], the force on the
Cu atom is computed in steps of $\frac{1}{72}$[111] for 13 points, then in
steps of $\frac{1}{24}$[111] until [111], where the DFT force dropped to
zero.  The Cu atom is directly above FCC and HCP sites to cancel forces in
the $(111)$ plane.  The difference in force between FCC and HCP is less
than 0.04eV/\AA\, for both DFT (max deviation at $\frac{1}{2}$[111]) and
EAM (max deviation at $\frac{2}{9}$[111]).  \Fig{111z} shows that DFT has a
stronger binding of Cu to the Ag surface than the Mishin EAM.  Also plotted
in \Fig{111z} is the force calculated with our optimized EAM, which
captures the deeper and wider well of DFT forces.

\bfig
\includegraphics[width=\smallfigwidth]{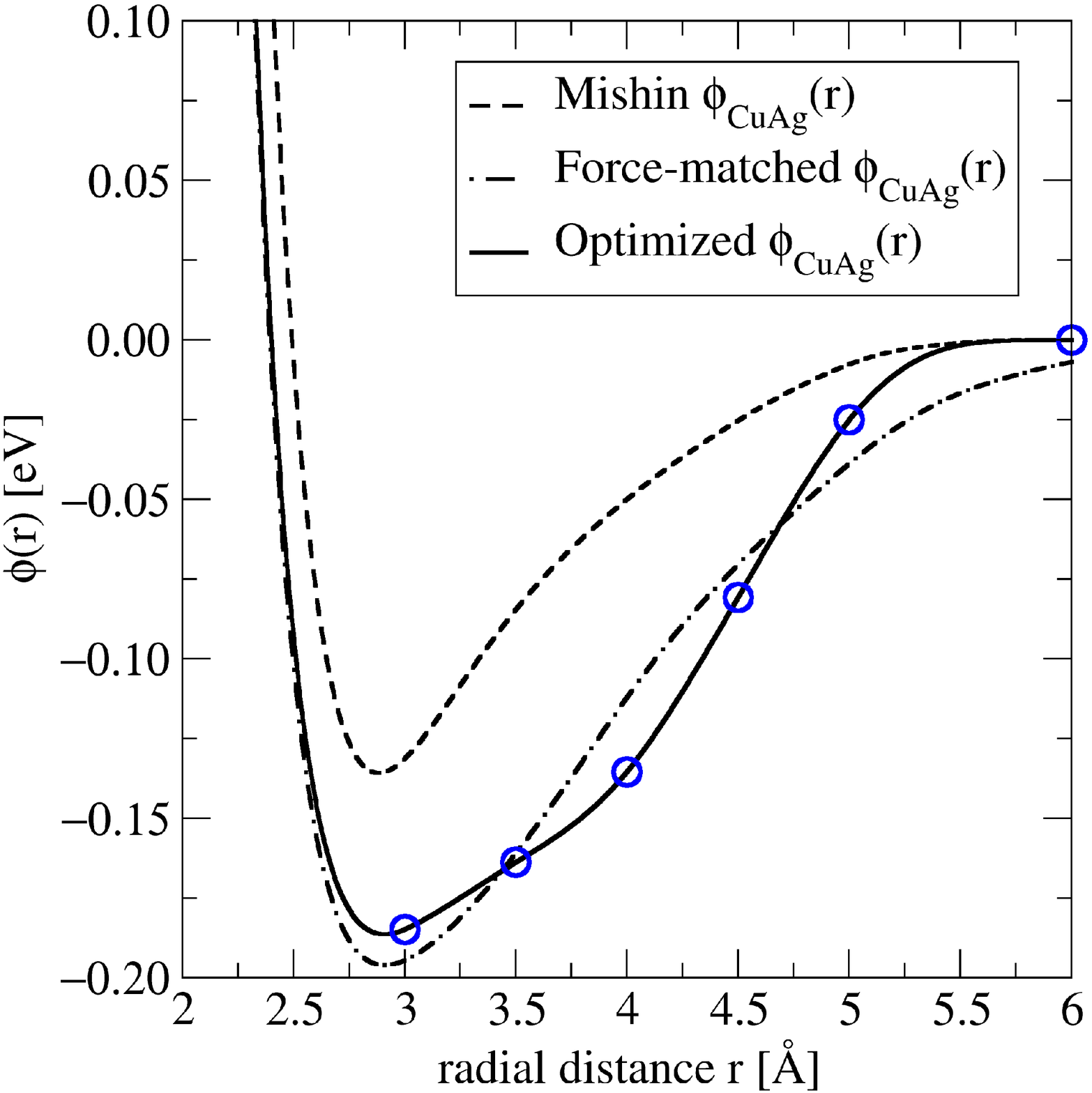}
\caption{The Cu-Ag pair potential at sequential steps in the optimization
process.  The integrated force difference from \Fig{111z} adds onto the
original Mishin \CuAg\ to produce the force-matched \CuAg.  Imposing a
smooth cut-off at 6\AA\ and adding a spline (knots at circles) to the
force-matched potential produces the optimized \CuAg.  The spline knots are
at 3\AA, 3.5\AA, 4\AA, 4.5\AA, 5\AA, and 6\AA.}
\label{fig:CuAg}
\efig

\Fig{CuAg} shows the Cu-Ag pair potential extracted from force-matching to
the DFT data.  Starting from the highest point and moving towards the
surface, the Cu atom feels the sum of forces from different shells of Ag
atoms within a 7.2\AA\ cut-off radius.  We chose this cutoff radius because
the DFT forces goes to zero at $z$=7\AA\ ([111]).  For each $z_{k}$, there
are $m$ = 1...$m_{k}$ shells, in which there are $n_{m}$ Ag atoms at
distance $r_{k,m}$ with directional component $c_{k,m} = \frac{\partial
r_{k,m}}{\partial z}\Bigl\vert_{z_{k}}$.  The $z$-component of the force at
height $z_{k}$ is
\be
\begin{split}
F_{z}(z_{k}) &= \frac{-\partial E_{\text{tot}}}{\partial z} = 
\sum_\text{all atoms} \frac{-\partial E_{\text{tot}}}{\partial r} \frac{\partial
r}{\partial z}\\
&= \sum_{m=1}^{m_{k}} -n_{m} c_{k,m}\phi'_{\text{CuAg}}(r_{k,m}),
\end{split}
\label{eqn:111z}
\ee
where $\phi'_{\text{CuAg}}(r_{k,m})$ is the radial derivative of the pair
potential.  We build the function $\phi'_{\text{CuAg}}(r)$ as a cubic
spline with knot points $r_{k}^{\text{knot}}$ = min\{$r_{k,m}$\} for each
$k$.  Starting from largest $z_{k}$ to smallest, \Eqn{111z} is solved for
$\phi'_{\text{CuAg}}(r_{k}^{\text{knot}})$ using, as needed, interpolated
values of $\phi'_{\text{CuAg}}(r)$ for $r > r_{k}^{\text{knot}}$.  The
equations are solved successively until $\phi'_{\text{CuAg}}$ is
self-consistent.  A final self-consistency loop over all FCC and HCP forces
is performed, alternating in sequence, obtaining $\phi'_{\text{CuAg}}(r)$
for $r$ in the range from 2.04\AA\ to 7.2\AA.  Integrating
$\phi'_{\text{CuAg}}(r)$ generates a quartic spline, the \CuAg\ plotted in
\Fig{CuAg}.  This force-matched \CuAg\ possess a deeper and wider energy well,
capturing the stronger Cu-Ag interaction from DFT.  For $r$ values smaller
than 2.04\AA, we linearly extrapolate \CuAg.

The force-matched \CuAg\ is refined by fitting to the monomer and dimer
database.  The force-matched \CuAg\ has inaccurate energies for monomers
and dimers, with the HCP site 4meV below the FCC site.  Modifying \CuCu\
does not affect monomer energies, and we find that the dimer energy
difference between FF and HH changes by less than 5meV with the \CuCu\
modifications we present later.  We optimize the Cu-Ag pair potential with
respect to monomer and homogeneous dimer site energy differences as the next
step.  We reduce the interaction range to 6\AA\ by shifting the potential
up by $\phi$(5.75\AA) and using quartic splines from 5\AA\ to 6\AA.  The
quartic splines have two equal spaced knots within the interval and matches
the value, first and second derivatives at 5\AA, and at 6\AA\ goes to zero
with zero slope and zero second derivative.  To differentiate between FCC
and HCP sites, we modify the Cu-Ag interaction in the range of the second
and third nearest neighbors for a Cu atom on the surface by adding a cubic
spline, with knots at 3.5\AA, 4\AA, 4.5\AA, 5\AA, and fixed end points at
3\AA\ and 6\AA.  We generate (2$\times$5+1)$^4$ = 14641 possible potentials
with different values at each knot point in steps of $\pm$20meV;
optimization continues using narrower ranges down to $\pm$1meV.  For each
sweep, we select potentials with the smallest rms monomer and homogeneous
dimer energy errors while also selecting for quantitatively low rms Cu
height errors and potentials without multiple minimums.  In \Fig{CuAg},
this optimized \CuAg\ exhibits a wider well than the force-matched pair
potential.

\Fig{CuCu} shows that the optimized \CuCu\ gives shorter and weaker
bonding between Cu atoms on the Ag surface than in the Mishin EAM bulk Cu.
We scale the original Mishin \CuCu\ in 1\% steps from 80\% to 120\%, and
translate in 0.01\AA\ steps from --0.15\AA\ to 0.15\AA; potentials with Cu
lattice parameter outside of $\pm$5\% of the bulk value are removed.  A
82\% scaling and a --0.13\AA\ translation reproduces all relative energy
differences with a final 0.5meV range optimization of the \CuAg.  We found
during optimization that although it was possible to obtain 0.012\AA\ rms
dimer separation error or 0.5meV rms energy error, these two errors grew
opposite one another.  We selected for lower energy error at the expense of
geometric agreement.

\bfig
\includegraphics[width=\smallfigwidth]{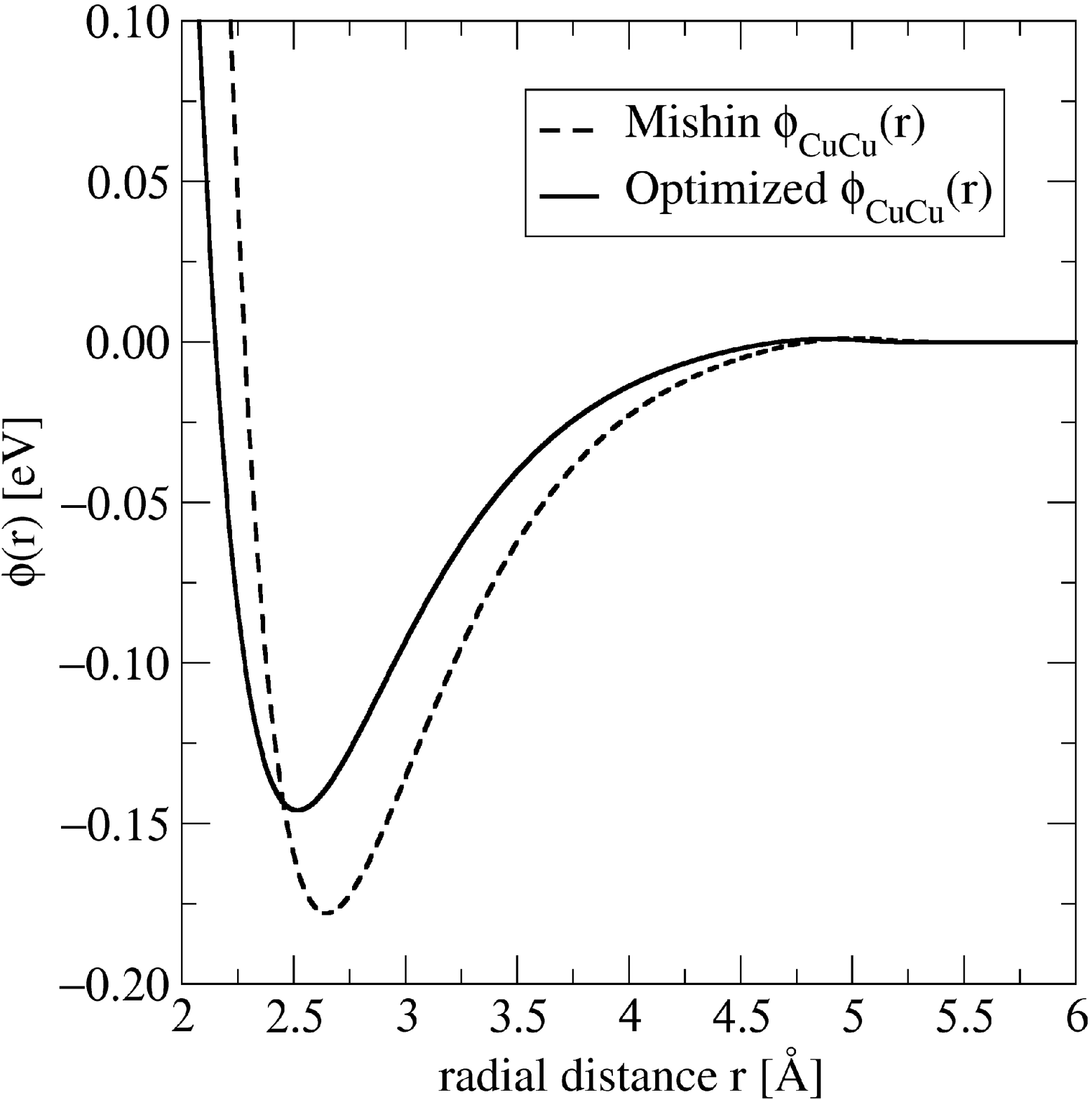}
\caption{The Cu-Cu pair potential from Mishin EAM and after
optimization. The original Mishin \CuCu\ is transformed with different
scalings and translations simultaneously with the optimization of the
\CuAg.  The optimized \CuCu\ is obtained from a scaling of 82\% and a
translation of --0.13\AA.  This indicates shorter and weaker bonding
between Cu atoms on the Ag surface than in Cu bulk.} 
\label{fig:CuCu}
\efig

Thus, our optimization procedure with respect to DFT Cu monomers and dimers
follows: (1) modify the \CuAg\ to match the force of an evaporating Cu atom
from Ag(111) calculated in DFT.  (2) Reduce the interaction length of the
force-matched \CuAg\ and add splines to reproduce the DFT FCC/HCP site
energy difference.  (3) Scale and translate the \CuCu\ to produce better
energetic agreement.  (4) Polish the optimized \CuAg\ potential with the
optimized \CuCu.  The final \CuAg\ is plotted in \Fig{CuAg}, and \CuCu\ in
\Fig{CuCu}.

\begin{table*}
\caption{Results for the optimized potential and other EAM potentials
compared to DFT and experimental values.  Three EAM potentials,
Foiles-Baskes-Daw, Voter-Chen, and Mishin were compared to the optimized
potential from this work with respect to monomer and dimer energies,
diffusion barriers and geometries.  Trimer energies for DFT and our
optimized potential are also presented.  The symbols indicate less than
20\% error or less than 0.1\AA $\; $error (\checkmark), between 20\% and
50\% error (---), and 50\% error or greater than 0.1\AA $\; $error
($\times$).  $\Delta$E(A,B) indicates the energy difference
E$_{\text{A}}$--E$_{\text{B}}$, and E$_{\text{a}}$(A$\rightarrow$B)
indicates the activation energy to transition from state A to state B.  The
\textit{ab initio} results, from ultrasoft pseudopotential DFT, are used as
the standard when available.  The optimized EAM potential in this work is
fit to DFT values indicated by the * symbol.}
\renewcommand{\arraystretch}{0.7} 
\label{tab:summary}
  \begin{ruledtabular}
\begin{tabular}{c cc c@{}c c@{}c c@{}c c@{}c}
 & 				& 			&\multicolumn{2}{c}{Foiles,}	&\multicolumn{2}{c}{ } 		&\multicolumn{2}{c}{ } 		&\multicolumn{2}{c}{ }  \\
 & 				& 			& \multicolumn{2}{c}{Baskes,}	& \multicolumn{2}{c}{Voter,} 	&\multicolumn{2}{c}{ } 		&\multicolumn{2}{c}{This} \\
 &Experiment\cite{Morgenstern:2004uq}	&\textit{ab initio} 	&\multicolumn{2}{c}{Daw\cite{Foiles:1986fj}} 	& \multicolumn{2}{c}{Chen\cite{Voter:1993kx}} 	&\multicolumn{2}{c}{Mishin\cite{Mishin:2001ym,Williams:2006qy}} 	&\multicolumn{2}{c}{Work} \\ \hline

& \multicolumn{10}{c}{\hrulefill \hspace{2mm} {\bf monomer energies [meV]} \hspace{2mm} \hrulefill} \\

$\Delta$E(H,F) 				& $ 5.5\pm 1.0$ & 14 	&  1  &$\times$	& 0&$\times$	& 8  &---		&  12 &\checkmark*  \\
E$_{\text{a}}$(F$\rightarrow$H)	& $65\pm 9$ 	&  	96	&   68  &$\times$& 39 &$\times$& 62 &$\times$& 93 &\checkmark  \\

& \multicolumn{10}{c}{\hrulefill \hspace{2mm} {\bf dimer energies [meV]} \hspace{2mm} \hrulefill} \\

$\Delta$E(HH,FF)  				& 			& 27 		& 1 		&$\times$		&   0  &$\times$ 	& 15 & ---			& 27  &\checkmark* \\
$\Delta$E(FH$_{\text{short}}$,FF) 	& 			& 71	 	& 58 		&---			&   2 &$\times$ 		&  79 &\checkmark 	& 71 &\checkmark*  \\
$\Delta$E(FH$_{\text{long}}$,FF) 	& 			& 134	& 66.5 	&$\times$ 	& 57 &$\times$ 	&  61 &$\times$ 	& 137 &\checkmark*  \\
$\Delta$E(FH$_{\text{long}}$,FH$_{\text{short}}$) &	& 63		& 8.5 	&$\times$ 	& 55 &\checkmark 	&  --18 &$\times$ 	& 66 &\checkmark  \\
E$_{\text{a}}$(FF$\rightarrow$HH) 	&  73		 	& 		& 62  	&\checkmark 	&   6  & $\times$	&  69 &\checkmark 	& 88 &--- \\

& \multicolumn{10}{c}{\hrulefill \hspace{2mm} {\bf trimer energies [meV]} \hspace{2mm} \hrulefill} \\

$\Delta$E(F$^{3}_{\text{non}}$,F$^{3}_{\text{rot}}$)  	& & --16 &41 &$\times$ &9 &$\times$ &33 &$\times$ & 9  &$\times$ \\
$\Delta$E(H$^{3}_{\text{non}}$,F$^{3}_{\text{rot}}$) 	& & 17  &38 &$\times$ &9 &$\times$ &50 &$\times$ & 42 &$\times$ \\
$\Delta$E(H$^{3}_{\text{rot}}$,F$^{3}_{\text{rot}}$) 	& & 42  &  2 &$\times$ &1 &$\times$ &23 &$\times$ & 45 &\checkmark  \\
$\Delta$E(H$^{3}_{\text{non}}$,F$^{3}_{\text{non}}$) & & 33  &--3 &$\times$ &0 &$\times$ &17 &$\times$ & 33 &\checkmark  \\

& \multicolumn{10}{c}{\hrulefill \hspace{2mm} {\bf geometries [\AA]} \hspace{2mm} \hrulefill} \\

Dimer length					&  &  \multirow{2}{*}{baseline}  	&\multirow{2}{*}{0.036}&\multirow{2}{*}{\checkmark}&\multirow{2}{*}{0.117}&\multirow{2}{*}{$\times$}&   \multirow{2}{*}{0.054} 	&\multirow{2}{*}{\checkmark} 	&   \multirow{2}{*}{0.0796}&\multirow{2}{*}{\checkmark*} \\ 
rms error						&  &   				&  &  &  &  &  					& 						&  					& \\
Adatom height 					&  &  \multirow{2}{*}{baseline} 	&\multirow{2}{*}{0.115}&\multirow{2}{*}{$\times$}&\multirow{2}{*}{0.559}&\multirow{2}{*}{$\times$}&   \multirow{2}{*}{0.153}	&\multirow{2}{*}{$\times$} 	&   \multirow{2}{*}{0.0397}&\multirow{2}{*}{\checkmark*}  \\
rms error  						&  &  					&  &  &  &  &  					&						& 					& \\
\end{tabular}
  \end{ruledtabular}
\end{table*}

In \Tab{summary}, comparison with other EAM potentials show that the
optimized EAM from this work has better agreement to DFT calculations.
Among the earlier potentials, the Mishin EAM%
\cite{Mishin:2001ym,Williams:2006qy} comes closest to the DFT energies
when compared to the Foiles-Baskes-Daw (FBD) EAM\cite{Foiles:1986fj} and
the Voter-Chen (VC) EAM\cite{Voter:1993kx} potentials.  The FBD and VC EAM
potentials did not indicate any site energy difference between FCC and HCP.
None of the earlier potentials were able to capture the correct DFT energy
difference between FH$_{\text{short}}$ and FH$_{\text{long}}$ dimers.  Both
FBD and Mishin EAM came within 10meV of the experimental monomer and dimer
diffusion barriers, while the optimized EAM potential overestimates the
experimental barriers but correctly predicts the DFT monomer diffusion
barrier.  The increased Cu-Ag interaction of the optimized potential were
able to pull the Cu atoms closer to the Ag surface, reducing the rms height
error over other potentials.  While the earlier potentials do not come
close to the DFT trimer energies, the optimized EAM is able to capture the
correct energies for $\Delta$E(H$^{3}_{\text{rot}}$,F$^{3}_{\text{rot}}$)
and $\Delta$E(H$^{3}_{\text{non}}$,F$^{3}_{\text{non}}$).  The deviation
for the trimer ground state will be discussed in \Sec{discussion}.

\section{Results}
\label{sec:results}

\bfig
\includegraphics[width=\smallfigwidth]{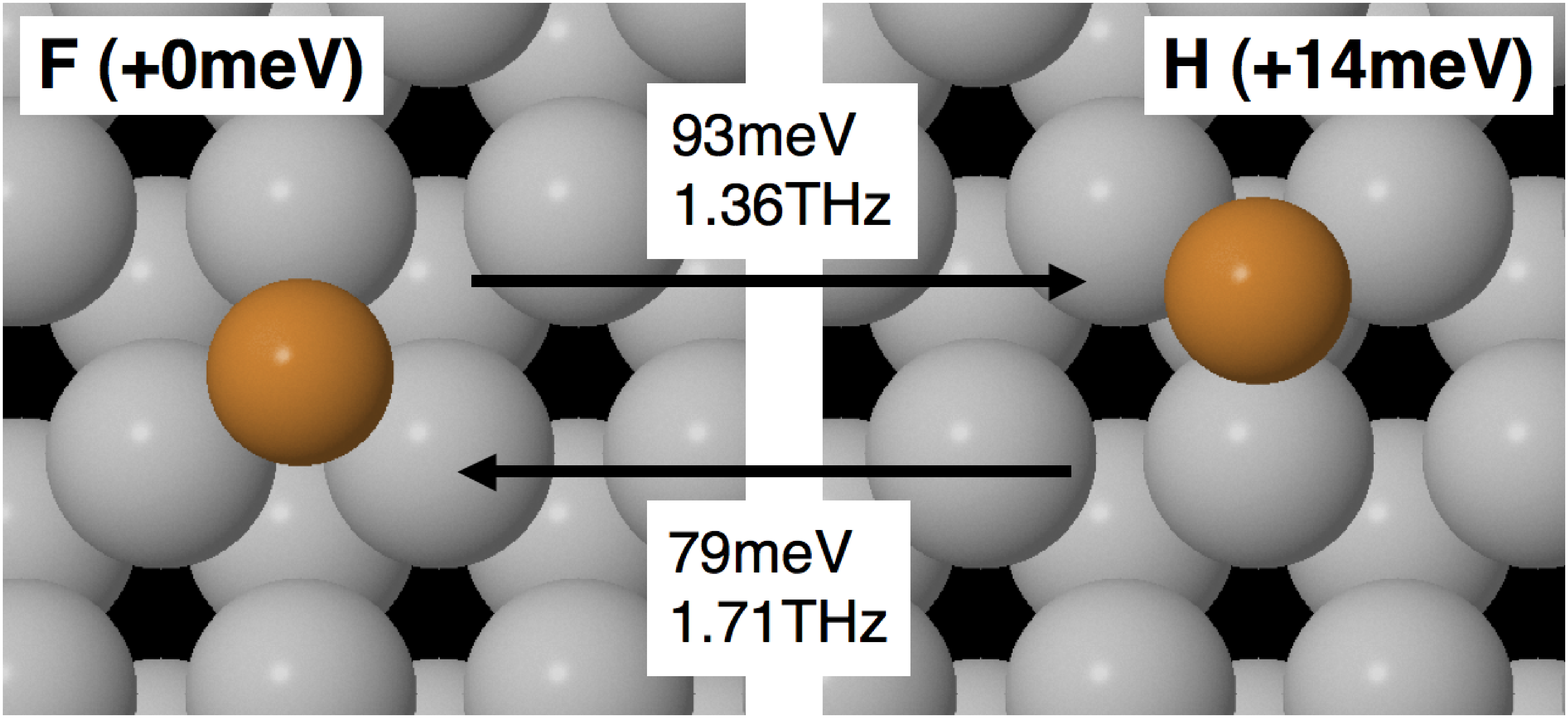}
\caption{Cu monomers FCC (F) and HCP (H), relative energy differences,
transition energies, and attempt frequency prefactors.  The FCC site is the
ground state and can diffuse to one of three equivalent HCP sites nearby;
similarly the HCP site diffuses to one of three FCC sites.  The
rate-limiting step in the diffusion process is the F$\rightarrow$H
transition with an energy barrier of 93meV.}
\label{fig:monomer}
\efig

\Fig{monomer} shows the geometries, relative energies, and transition
barriers of Cu monomers calculated with the optimized EAM.  The 14meV
energy difference between the FCC and HCP site also represents the
difference between the transition barriers.  The two transitions possible
are the F$\rightarrow$H with a 93meV barrier and the 79meV barrier
H$\rightarrow$F transition.  The F$\rightarrow$H barrier is higher than the
experimental value of 65$\pm$9meV,\cite{Morgenstern:2004uq} but matches our
DFT calculations for the bridging site with an energy difference of 96meV.
DFT is known to overestimate surface adsorption,\cite{Stampfl:2005lr} and
we discuss strategies to compensate in \Sec{discussion}.  The agreement
with DFT is a confirmation of our potential since the bridging site energy
is not part of the optimization database.

\bfig
\includegraphics[width=\smallfigwidth]{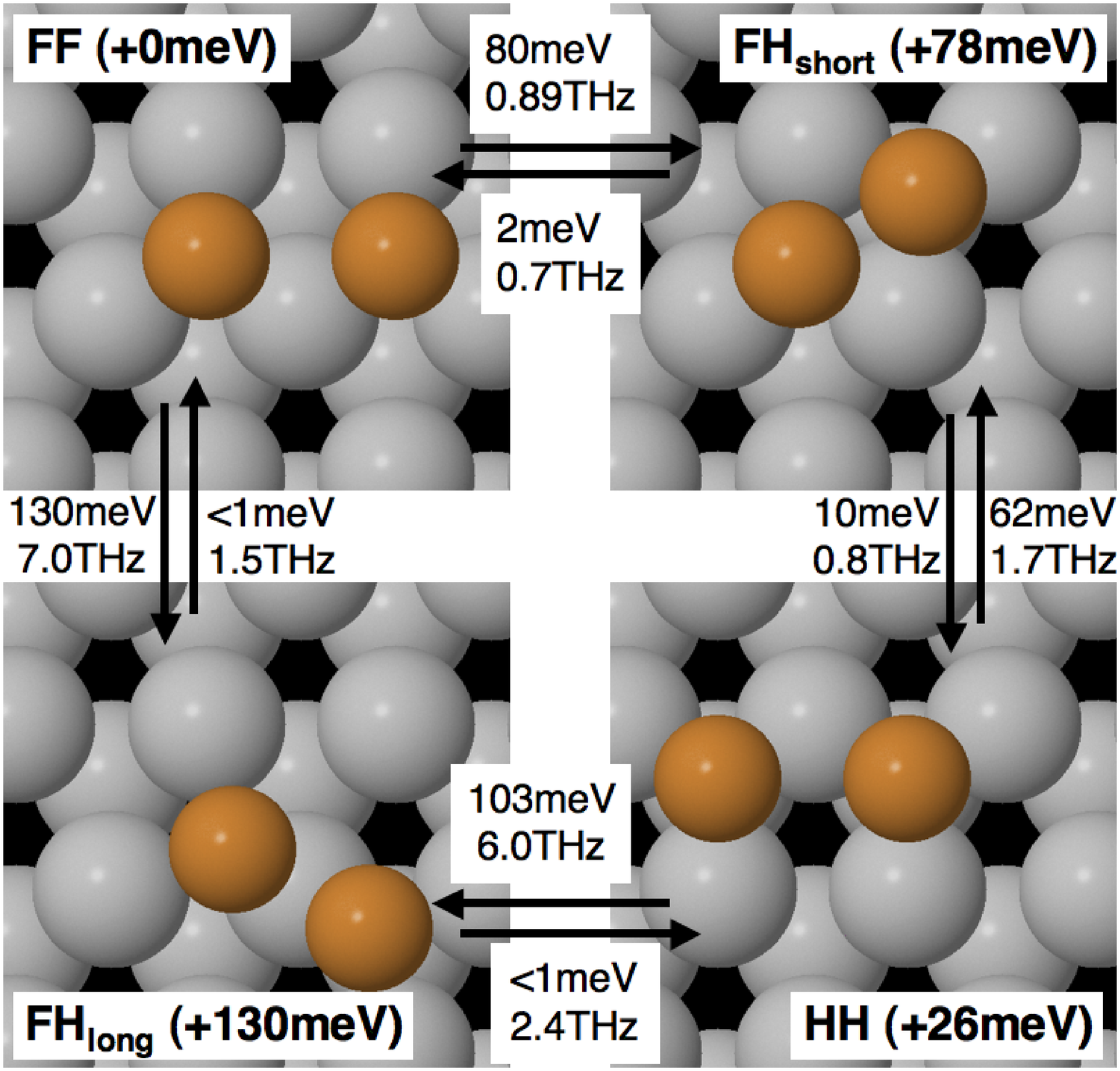}
\caption{Cu dimers FCC-FCC (FF), FCC-HCP neighboring (FH$_{\text{short}}$),
FCC-HCP non-neighboring (FH$_{\text{long}}$), and HCP-HCP (HH), relative
energy differences, transition energies, and attempt frequency prefactors.
The FF dimer is the ground state and diffusion to the HH site is achieved
through one of two FH meta-stable sites.  The pathway through the
FH$_{\text{short}}$ dominates the FF$\leftrightarrow$HH diffusion, giving
an overall rate-limiting barrier of 88meV
(FF$\rightarrow$FH$_{\text{short}}$$\rightarrow$HH, 80meV -- 2meV +
10meV).}
\label{fig:dimer}
\efig

\Fig{dimer} shows the geometries, relative energies, and transition
barriers of Cu dimers calculated with the optimized EAM.  The FF dimer is
the ground state and the HH dimer is 26meV higher in energy, about twice
the monomer energy difference.  The FH$_{\text{short}}$ and
FH$_{\text{long}}$ dimers are two metastable configurations which are 78meV
and 130meV higher in energy than FF, respectively.  Dimer diffusion is more
complex than that for monomers, with two intermediate states between FF and
HH plus dimer rotation.  With low barrier ($<$1meV) transitions out of the
FH$_{\text{long}}$ state, the diffusion pathway through FH$_{\text{long}}$
has a 130meV barrier for FF$\rightarrow$FH$_{\text{long}}$$\rightarrow$HH,
and 103meV barrier for HH$\rightarrow$FH$_{\text{long}}$$\rightarrow$FF.
The other diffusion pathway is more complicated, since an
FH$_{\text{short}}$ dimer is more likely to transition to FF (2meV barrier)
than to HH (10meV barrier).  This results in a 88meV barrier for
FF$\rightarrow$FH$_{\text{short}}$$\rightarrow$HH (88meV -- 2meV + 10meV),
and 62meV barrier for HH$\rightarrow$FH$_{\text{short}}$$\rightarrow$FF.
The calculated barriers are higher than the experimental barrier of
73meV,\cite{Morgenstern:2004uq} again consistent with overestimated
adsorption energies by DFT.

\bfig
\includegraphics[width=\smallfigwidth]{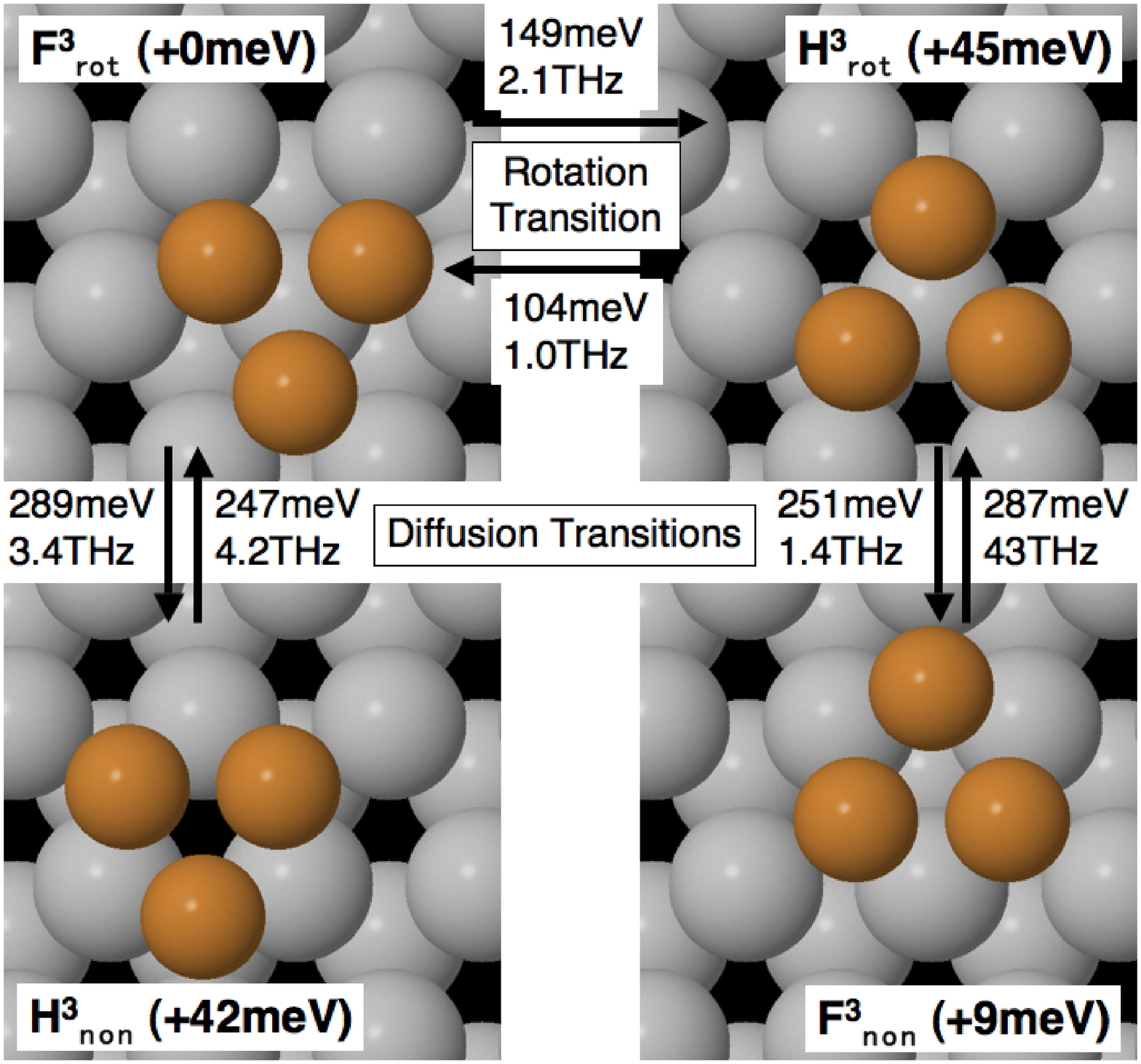}
\caption{Cu trimers FFF centered on Ag (F$^3_\text{rot}$), HHH centered on
Ag (H$^3_\text{rot}$), HHH centered on a hole (H$^3_\text{non}$), and FFF
centered on a hole (F$^3_\text{non}$), relative energy differences,
transition energies, and attempt frequency prefactors.  F$^3_\text{rot}$ is
the ground state but is only 9meV lower in energy than the F$^3_\text{non}$
state.  A rotation transition exists between F$^3_\text{rot}$ and
H$^3_\text{rot}$, while the system can diffuse by overcoming the higher
289meV and 287meV energy barriers for
F$^3_\text{rot}$$\rightarrow$H$^3_\text{non}$ and
H$^3_\text{rot}$$\rightarrow$F$^3_\text{non}$ respectively.}
\label{fig:trimer}
\efig

\Fig{trimer} shows the geometries, relative energies, and transition
barriers of Cu trimers calculated with the optimized EAM.  There are two
different configurations for each of the FCC and HCP trimers due to the
geometry of the (111) surface.  The trimer triangles can either be centered
around a surface Ag atom permitting rotation, F$^{3}_{\text{rot}}$
(groundstate) and H$^{3}_{\text{rot}}$ (+45meV), or not,
F$^{3}_{\text{non}}$ (+9meV) and H$^{3}_{\text{non}}$ (+42meV).  The
relative energy difference between F$^{3}$ and H$^{3}$ trimers is
approximately two to three times the monomer energy difference.  The
rotation transition F$^{3}_{\text{rot}}$$\rightarrow$H$^{3}_{\text{rot}}$
has a 149meV barrier and a 104meV barrier for the reverse.  The non trimers
do not rotate, and transition to rot trimers on the opposite sites.  These
transition barriers are higher than the rotation barriers, at $\sim$290meV
from F$^{3}_{\text{non}}$$\rightarrow$H$^{3}_{\text{rot}}$ and
F$^{3}_{\text{rot}}$$\rightarrow$H$^{3}_{\text{non}}$, and $\sim$250meV for
the reverse.  We expect, as with monomers and dimers, that the EAM
overestimates the trimer transition barriers.

We construct analytical expressions for the diffusion constants of
monomers, dimers, and trimers using the calculated transition barriers and
attempt frequencies.  The rate of jumping from a F site to a particular H
site is $r_{\text{FH}} = \nu_{\text{F}\rightarrow \text{H}}
\exp\left(-\text{E}_{\text{a}}(\text{F}\rightarrow \text{H})/
k_\text{B}T\right)$ where $\text{E}_{\text{a}}(\text{F}\rightarrow
\text{H})$ and $\nu_{\text{F}\rightarrow \text{H}}$ are the energy barrier
and the attempt frequency for the F to H transition.  Then a monomer
moving from one F to a new F site through an H site at temperature $T$
occurs with mean wait time of
\beu
\tau_{\text{monomer}} = \frac{3}{2} \left[ (3 r_{\text{FH}})^{-1} + (3
r_{\text{HF}})^{-1}  \right],
\eeu
%
including the three equivalent hopping sites for each monomer transition,
and with a correlation factor of $\frac{3}{2}$ for monomer transitions to
the original site.  The Einstein diffusion relation, $D =
\frac{1}{4}a_{nn}^2 \tau^{-1}$ where $a_{nn}=2.89\text{\AA}$ is the
nearest-neighbor distance between Ag atoms gives the monomer diffusion
constant
\beu
D_{\text{monomer}} = \frac{a_{nn}^2}{2} \left[ (r_{\text{FH}})^{-1} +
(r_{\text{HF}})^{-1}\right]^{-1}.
\eeu
For both the dimer and trimer case, the diffusion system becomes complex
and we use the continuous-time random walk formalism developed by
Shlesinger and Landman.\cite{Shlesinger:1976lr}  The diffusion constant for
the dimer is computed numerically and plotted in \Fig{diffusion}, while the
diffusion constant for the trimer is given by
\begin{widetext}
\beu
D_{\text{trimer}} = \frac{a_{nn}^{2}}{2}
\frac{r_{\text{F$^{3}_{\text{non}}$H$^{3}_{\text{rot}}$}}
r_{\text{H$^{3}_{\text{non}}$F$^{3}_{\text{rot}}$}}
(r_{\text{F$^{3}_{\text{rot}}$H$^{3}_{\text{rot}}$}}
 r_{\text{H$^{3}_{\text{rot}}$F$^{3}_{\text{non}}$}}
+r_{\text{F$^{3}_{\text{rot}}$H$^{3}_{\text{non}}$}}
r_{\text{H$^{3}_{\text{rot}}$F$^{3}_{\text{rot}}$}})}%
{r_{\text{F$^{3}_{\text{rot}}$H$^{3}_{\text{rot}}$}}
 r_{\text{H$^{3}_{\text{non}}$F$^{3}_{\text{rot}}$}}
 r_{\text{H$^{3}_{\text{rot}}$F$^{3}_{\text{non}}$}}
+r_{\text{F$^{3}_{\text{non}}$H$^{3}_{\text{rot}}$}}
(r_{\text{F$^{3}_{\text{rot}}$H$^{3}_{\text{rot}}$}}
r_{\text{H$^{3}_{\text{non}}$F$^{3}_{\text{rot}}$}}
+(r_{\text{F$^{3}_{\text{rot}}$H$^{3}_{\text{non}}$}}
+r_{\text{H$^{3}_{\text{non}}$F$^{3}_{\text{rot}}$}})
 r_{\text{H$^{3}_{\text{rot}}$F$^{3}_{\text{rot}}$}})}.
\eeu
\end{widetext}
%

In \Fig{diffusion}, the analytical rates from above have been plotted as
diffusion coefficients against temperature along with experimental data
from [\onlinecite{Morgenstern:2004uq}] for the monomer and dimer.  The
experimental barriers, 65$\pm$9meV and 73meV for monomer and dimer are both
lower than our calculated values, though no error bar is given for the
dimer experimental barrier.  The rate limiting barriers as T$\rightarrow$0K
are calculated using data at T$<$20K.  The rate limiting barriers, 93meV,
88meV, and 289meV, correspond to the rate-limiting transition barriers
identified above for the monomer, dimer, and trimer.  The dimer diffusion
slope decreases with increasing temperature due to the influence of both FH
intermediate states.  Higher temperatures samples the FH$_{\text{long}}$
pathway; this decreases dimer diffusion as transitions through the
FH$_{\text{long}}$ state lead to rotation, i.e.,
FF$\rightarrow$FH$_{\text{long}}$$\rightarrow$HH$\rightarrow$FH$_{\text{long}}$$\rightarrow$FF.
The prefactors for transitions out of FH$_{\text{short}}$ are the lowest
for all dimer transitions, and become rate limiting at high temperatures.

\bfig
\includegraphics[width=\smallfigwidth]{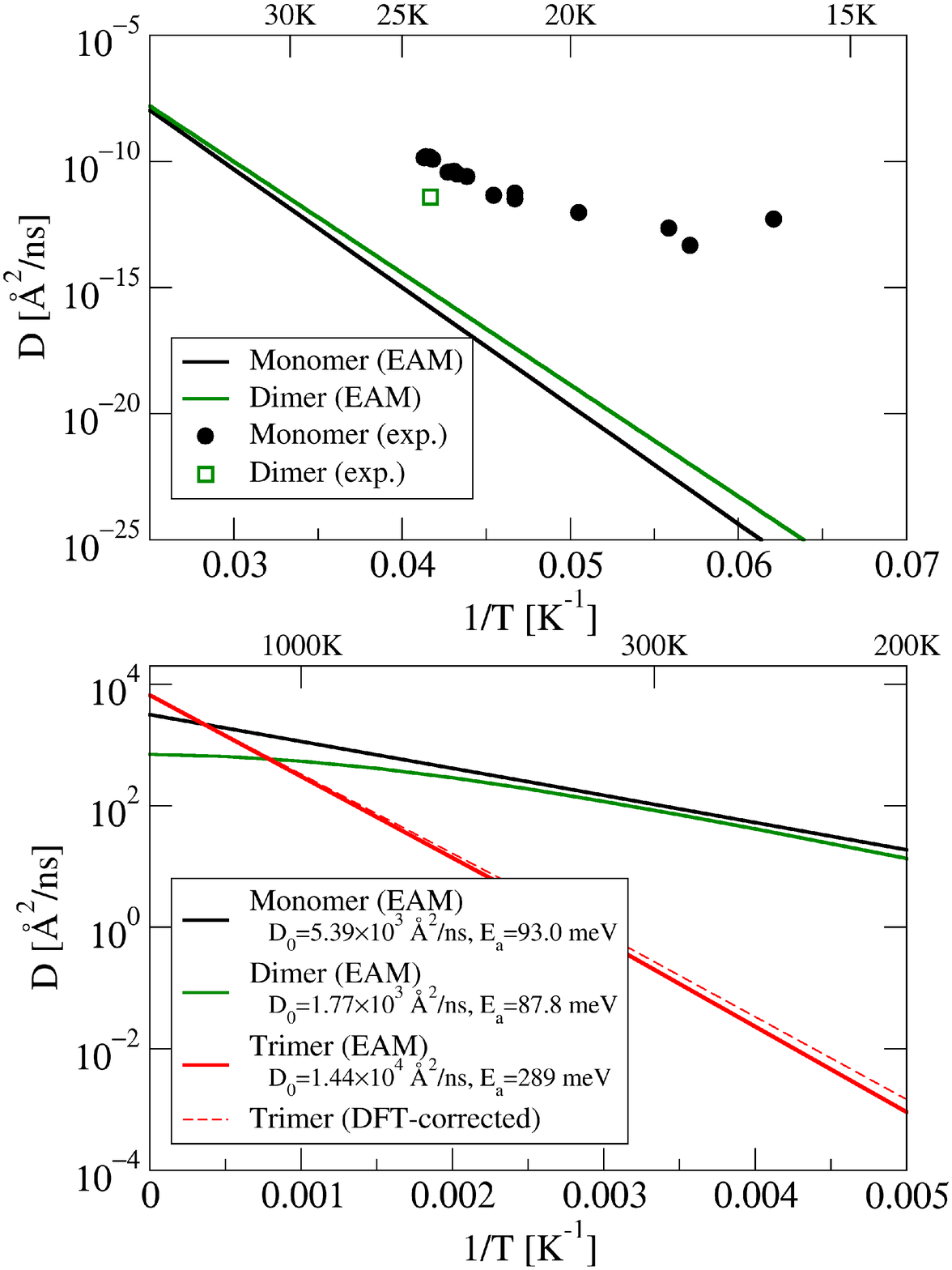}
\caption{Experimental results for monomer and dimer compared with
analytical diffusion calculations for monomer, dimer, trimer, and DFT
corrected trimer at low (top) and high (bottom) temperatures.  The
Arrhenius fit, in the T$\rightarrow$0K limit, for the monomer, dimer, and
trimer systems reflects the rate-limiting diffusion barriers 93meV, 88meV,
and 289meV respectively.  The DFT corrected trimer is calculated by
adjusting the trimer diffusion barriers to match DFT energy differences and
keeping the same prefactors.  The experimental monomer barrier is
65$\pm$9meV from data in the temperature range 19K--25K.  The experimental
dimer barrier is 73meV from data at 24K assuming a prefactor of 1THz.}
\label{fig:diffusion}
\efig

\section{Discussion}
\label{sec:discussion}
The Cu monomer is the basic unit for Cu islands on Ag(111), and the correct
extrapolation of monomer energies and barriers to dimers and trimers
indicates the optimized potential is consistent with DFT.  For the monomer,
the site energy difference between FCC and HCP is 14meV, twice the
difference is seen between the homogeneous dimers FF and HH at 26meV, and
two to three times the difference in the rot and non trimer pairs at 33meV
and 45meV.  The diffusion barriers for the trimer are also three times that
of the monomer 289meV versus 93meV and 247meV versus 79meV.  This linear
relationship is explained by the fact that in the trimer diffusional
transitions, all three atoms move simultaneously over each of their
respective bridging sites, thus the trimer as a whole experiences a barrier
three times as large.  In the dimer system, diffusion moves one atom at a
time and the barrier is comparable to that of the monomer.

EAM produces higher diffusion barriers for monomer and dimer than in
experiment,\cite{Morgenstern:2004uq} but gives diffusion barriers that
match DFT.  This effect is consistent with the observed overestimation of
surface adsorption energy by DFT calculations.\cite{Stampfl:2005lr}
Compared to experiments for monomer and dimer, the barriers are
overestimated by approximately 10 to 15meV.  Since diffusion for both the
monomer and dimer proceeds one Cu atom at a time, we expect the bridging
site between F and H to be overestimated by 15meV.  For general diffusion
barriers, a 15meV reduction should be applied for each concurrent Cu atom
in the transition when comparing to experiment.  For example, a three-fold
reduction of 45meV will need to be applied to the trimer diffusion
barriers.

Trimers were not included in our optimization database and calculations in
DFT and the optimized EAM differ when looking at the energetics between rot
and non trimers.  DFT calculates that the two non trimers are 25meV lower
in energy than predicted by our optimized EAM, making the ground state
trimer configuration F$^{3}_{\text{non}}$ rather than F$^{3}_{\text{rot}}$.
We expect the deviation to be mainly caused by the center Ag atom under the
rot trimer, whose embedded ``electron density'' is 16\% higher than an Ag
atom in the bulk.  This density is beyond the range present in the monomer
and dimer database.  Modifying the embedding function, as done in surface
embedded-atom method (SEAM)\cite{Haftel:1993lr,Haftel:1995fk} may offset
this effect by penalizing densities away from the bulk value.  Although the
relative energy between rot and non trimers are not correct, the optimized
EAM correctly predicts the energy difference between F$^{3}_{\text{rot}}$
and H$^{3}_{\text{rot}}$, and F$^{3}_{\text{non}}$ and
H$^{3}_{\text{non}}$.  Adding a Cu atom to a rot trimer will create a non
trimer subsection, and in larger islands, this pairing of rot and non
trimers allows the correct energy differences to be calculated.  We expect
the trimer diffusion barrier to remain three times that of the monomer even
with the change in ground state.  A new estimate of trimer diffusion can be
computed by splitting the 25meV energy difference between forward and
reverse diffusion barriers, e.g. lower the rot to non barrier by 12.5meV
and raising the non to rot by 12.5meV.  This change does not affect the
transition paths and therefore does not change the overall diffusional
dynamics of the trimer system, increasing the rate limiting barrier to
292meV from 289meV (c.f.~\Fig{diffusion}).  Applying the 45meV
over-adsorption correction gives a barrier of 247meV for trimer diffusion
to compare with experiment.

\section{Conclusion}
\label{sec:concl}
We present a method to optimize an EAM potential for heterogeneous surface
system using {\it ab initio} data.  The optimized EAM potential reproduces
DFT energies for Cu monomers, dimers, and trimers on Ag(111).  Diffusion
barriers for monomers, dimers, and trimers are calculated to be 93meV,
88meV, and 289meV, which match available DFT data, but exceed experimental
values.  To correct for the overestimated barriers, a 15meV reduction is
applied for each concurrently transitioning Cu atom.  We found a 25meV
energy discrepancy between rot and non trimers when compared with DFT.
This discrepancy is not worse for larger islands due to correct energy
difference between F-trimers and H-trimers calculated by the potential
compared with DFT.  We expect the new EAM potential to accurately describe
the diffusion and energetics of larger Cu islands on Ag(111).

\begin{acknowledgments}
The authors thank John Weaver and Andrew Signor for helpful discussions.
This research was supported by NSF/DMR grant 0703995, and 3M's Untenured
Faculty Research Award.
\end{acknowledgments}

\end{document}